\documentclass[12pt]{iopart}
\usepackage{epsf,iopams}
\begin{document}
\title{Spin-orbit splitting of image states}
\author{J R McLaughlan, E M Llewellyn-Samuel and 
S Crampin\footnote[1]{To whom correspondance should be addressed (s.crampin@bath.ac.uk)}}
\address{Department of Physics, University of Bath, Bath, BA2 7AY, United Kingdom}
\begin{abstract}
We quantify the effect of the spin-orbit interaction on the Rydberg-like series of
image state electrons 
at the (111) and (001) surface of Ir, Pt and Au. Using relativistic multiple-scattering
methods we find Rashba-like dispersions with $\Delta E^{\rm SO}(K)=\gamma K$ 
with
values of $\gamma$ for $n=1$ states in the range $38-88~\rm{meV}~\mathrm{\AA}$. Extending the 
phase-accumulation model to include spin-orbit scattering we find that the splittings
vary like $1/(n+a)^3$ where $a$ is the quantum defect and that they are related to the
probability of spin-flip scattering at the surface. The splittings should be observable
experimentally
being larger in magnitude than some exchange-splittings that have been resolved by
inverse photoemission,
and are comparable to linewidths from inelastic lifetimes.
\end{abstract}
\submitto{\JPCM}

\section{Introduction}

Image states \cite{echenique78}
are a special class of weakly bound surface electronic states 
in which an electron outside a dielectric or conductor polarises the surface
and is then attracted to the resulting ``image charge''. Asymptotically 
the potential varies like $V(\bi{r})\sim -(4z)^{-1}$ so that a band gap 
preventing
penetration of electrons into the crystal leads to a Rydberg-like series of 
states which in the case of a purely Coulombic image force at a planar 
metal surface arise at energies 
$E_n(\bi{K})=-[0.85~\mathrm{eV}]/n^2+\hbar^2K^2/2m$,
$n=1,2\dots$
where $\bi{K}$ is the electron wave vector parallel to the surface 
and $m$ the electron mass.
Deviations from this behaviour reflect the influence and response 
of the surface-dependant 
electronic and atomic structure, which may therefore be investigated by 
studying image states. 
Examples of theoretical and experimental work
include the systematics of image states binding and dispersion
on clean surfaces \cite{straub86}, 
image state on overlayers \cite{padowitz92,fischer93a,wallauer96},
at stepped metal surfaces \cite{smadici04}, 
exchange splitting of image states at 
ferromagnets \cite{passak92,nekovee93,derossi96}, 
as well as image states at surface nanostructures 
\cite{fischer93b,ortega94,hill99,kasperovich00}.
In recent years there has also been considerable interest in the dynamics
of image state electron \cite{echenique00,wahl03,rhie03,boger04}
as model electronic excitations at surfaces.

One aspect of the physics of image state electrons that has yet to be 
addressed is the influence of the spin orbit interaction 
$H^{\rm SO}=(\hbar/4m^2c^2)\bsigma\cdot\left(\bnabla V_{\textstyle\times} \bi{p}\right)$
\cite{messiah62}
which has recently
been found to have a significant effect on other surface state electron levels
at the surfaces of conductors with high atomic 
number \cite{lashell96,rotenberg99}. At first sight the spin-orbit
interaction might be expected to be negligible. The mathematical analogy 
that can be drawn between the Schr\"odinger equation describing the
electrons moving in the Coulomb-like image potential and that of $s$-electrons
in the hydrogen atom enable the image state wavefunctions to be written as
\begin{equation}
\psi_{n,\bi{K},s}(\bi{r})=(1/8)zR_{n0}(z/4)\exp(\rmi\bi{K}\cdot\bi{r}_\|)\chi_s
\end{equation}
where $R_{n\ell}(r)$ is the normalised radial hydrogenic wavefunction and
$\chi_s$ a Pauli spinor: 
$\chi_\uparrow={1\choose 0}, \chi_\downarrow={0 \choose 1}$.
Using these wavefunctions to diagonalise the spin-orbit perturbation 
$H^{\rm SO}$ in the subspace of degenerate
image state levels gives a spin-orbit splitting of
\begin{equation}
\Delta E_n^{\rm SO}=\frac{\alpha^2 e^2 K}{64(4\pi\epsilon_0)n^3}=
\left[0.012~\rm{meV}~\mathrm{\AA}\right]\frac{K}{n^3}
\end{equation}
where $\alpha=e^2/4\pi\epsilon_0\hbar c$ is the fine structure constant. 
This is well below the current resolution of inverse photoemission,
two-photon photoemission or scanning tunnelling 
spectroscopy. However, it has previously been 
recognised that a more significant contribution to the spin-orbit splitting 
of ``crystal--derived'' surface states arises from the brief time spent by 
the electron in the vicinity of the nuclei of the surface atoms, where the 
gradient contribution to the spin-orbit interaction is 
$|\bnabla V|\sim Z/r^2$.  In this paper we report on
calculations that we have performed to quantify the
magnitude of the spin-orbit splitting that arises from the
penetration of the image state wave function into the crystal at surfaces
of Ir, Pt and Au. These are described in section \ref{section:rlkkr}. In
section \ref{section:phase} we describe the modification of the 
phase accumulation model for image state energetics to include the
effects of the spin orbit interaction. Finally, we summarise and discuss 
our findings.

\section{Relativistic electronic structure calculations}
\label{section:rlkkr}

To calculate the spin-orbit splitting of image states we use a recently
developed code that implements relativistic multiple-scattering theory.
The theory behind this method is essentially that described by Halilov
\textit{et al.}
\cite{halilov93} so we do not reproduce it in detail here. The basic idea is that
the electronic structure is found from the single-particle Green function 
corresponding to the Dirac Hamiltonian $\hat{H}=c\bi{\alpha}\cdot\bi{p}+\beta mc^2+V$
\cite{messiah62}. Thus spin-orbit effects are treated non-perturbatively.
Using scattering techniques the Green function is determined for the special case of
semi-infinite crystals with two-dimensional in-plane translational periodicity,
treating intralayer scattering within an angular momentum representation and
interlayer scattering in a plane wave representation.
Our calculations use 25 and 19 two-dimensional reciprocal lattice vectors 
to describe the interlayer scattering for the (001) and (111) surfaces 
respectively, and partial waves up to $\ell_{\rm max}=4$ \cite{crampin94}. 
The semi-infinite substrate means that continuum and surface-localised states 
are clearly distinguished in the
wave vector resolved local density of states, found from the imaginary part of the
Green function.  As in the non-relativistic version of the code \cite{crampin93}
the electronic structure is found self-consistently using the 
local density approximation to density functional theory.
We use the atomic sphere approximation for the crystal potential 
(including dipole contributions), with the potential in the three outermost
atomic layers allowed to vary in response to the presence of the surface.
Since the local density approximation does not lead to an image-like surface barrier, 
and hence does not support image states,
once self-consistency has been achieved we replace the self-consistent barrier 
with a parameterised model barrier, for which we use the ``JJJ'' potential 
\cite{jones84}
\begin{equation}
V_{\rm B}(z)=\left\{
\begin{array}{l}
\left(1-e^{\lambda(z-z_0)}\right)/4(z-z_0),\quad z<z_0\\
-U/\left(1+Ae^{-\beta(z-z_0)}\right),\quad z>z_0.
\end{array}
\right.
\end{equation}
The fitting parameters $\lambda$, $U$ and $z_0$ were fixed by starting with values quoted by
Smith \textit{et al.} \cite{smith89}, who fitted to first-principles slab calculations, 
and then adjusted slightly to place the $n=1$ image state at
$K=0$, $E_1$, close to values found experimentally. The procedure does not uniquely fix the
parameters, but we found that different combinations that gave the same value for $E_1$
resulted in almost identical image state dispersion curves. Note that our results are for
the $1\times 1$ unreconstructed surfaces of the materials studied. In several
cases the surfaces undergo complex surface reconstructions (e.g. Au(001) and Pt(001) adopt a 
$5\times 20$ reconstruction but may be prepared in the $1\times 1$ structure
-- see \cite{drube88}).  

\begin{figure}
\epsfxsize= 160mm
\centerline{\epsffile{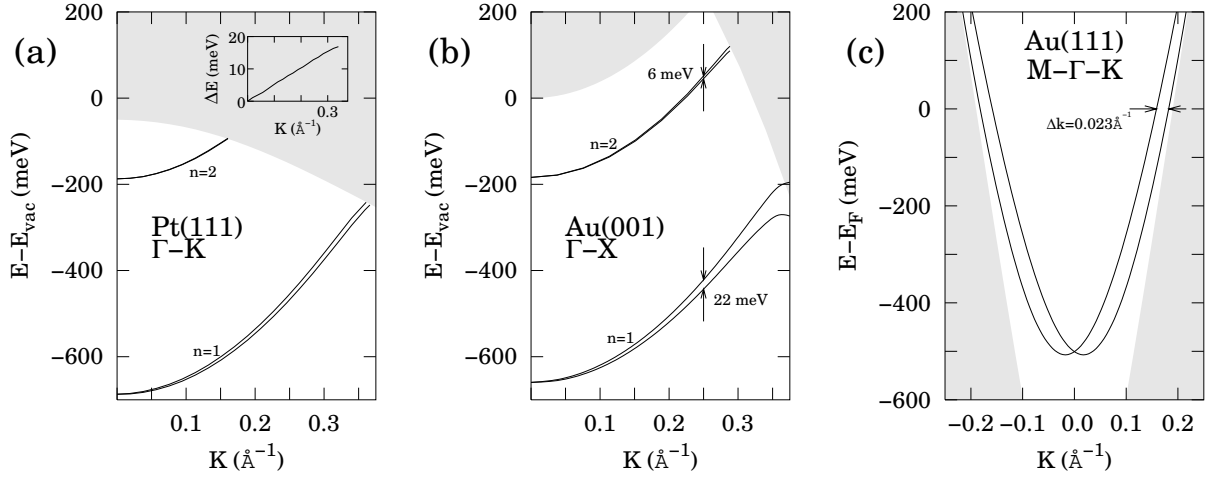}}
\caption{Calculated surface state dispersion curves. The shading indicates the presence of
bulk or vacuum continuum states. 
(a) Pt(111) $n=1$ and $n=2$
image states. The inset demonstrates the splitting of the $n=1$ state is linear in $K$.
(b) Au(001) $n=1$ and $n=2$ image states. 
(c) Au(111) surface state (note the energy scale in this case is with reference 
to the Fermi level).}
\label{fig:splits}
\end{figure}
In figure \ref{fig:splits} we illustrate the dispersion curves that we find.
For both Pt(111) and Au(001) the spin-orbit interaction can clearly be seen to split the 
$n=1$ image state, and whilst a splitting exists for the $n=2$ and higher states it is much 
smaller.  The inset in the figure \ref{fig:splits}a illustrates the variation
of the splitting with wave vector $K$, the near-linear variation corresponding to a Rashba-like
dispersion \cite{rashba60,bychkov84}
\begin{equation}
E_n(K) \simeq E_n+\frac{\hbar^2K^2}{2m}\pm (\gamma/2) K.
\end{equation}
From curves such as these we extract Rashba-parameters $\gamma$ by a least-square fit using
wave vectors $K \le 0.2~\mathrm{\AA}^{-1}$. These values are tabulated in
table \ref{table:results} for the image states at the 
(111) and (001) surfaces of Ir, Pt and Au. At Au(111) the vacuum level lies outside of
the projected band gap so that the image states in this case exist as  resonances. We 
have not therefore included results for this case, but instead give the results that we find
for the spin orbit
splitting of the occupied surface state that occurs within the band gap at this surface.
The dispersion of this state is shown also in figure \ref{fig:splits}, and agrees well
with previous work, validating our procedure. 
We find the wave vector splitting at the Fermi energy is $\Delta k=0.023~\mathrm{\AA}^{-1}$, 
compared with experimental values of $0.023~\mathrm{\AA}^{-1}$ \cite{lashell96}, 
$0.025~\mathrm{\AA}^{-1}$ \cite{nicolay01,reinert03} and $0.027~\mathrm{\AA}^{-1}$ \cite{mugarza02},
and a theoretical value of $0.023~\mathrm{\AA}^{-1}$ \cite{henk03} that have been 
reported previously.

\begin{table}
\caption{\label{table:results}Calculated image state energies including
the spin-orbit interaction at
$(001)-1\times 1$ and $(111)-1\times 1$ surfaces of Ir, Pt and Au: 
$E_n(K)=E_n+\hbar^2K^2/2m\pm(\gamma/2) K$. $E_n$ values in brackets are from experiment.}
\begin{indented}
\item[]
\begin{tabular}{ccccccc}
\br
surface&~~$n$~~&$E_n$ (eV)&$m(m_e)$ & $\gamma$ (meV~\AA)\\
\mr
Ir(111)&1& -0.65 &$0.95$&$56\pm 1$\\
       &2& -0.18&$1.00$&$\phantom{0}9\pm 1$\\
Ir(001)&1& -0.61&$0.94$&$38\pm 1$\\
       &2& -0.17&$0.99$&$\phantom{0}6\pm 1$\\
Pt(111)&1& -0.69 (-0.65$^a$, -0.78$^b$)&$1.05$&$50\pm 2$\\
       &2& -0.19 (-0.16$^a$, -0.20$^b$)&$1.03$&$\phantom{0}8\pm 1$\\
Pt(001)&1& -0.60 (-0.60$^c$)&$0.96$&$47\pm 2$\\
       &2& -0.17 &$0.99$&$\phantom{0}9\pm 1$\\
Au(111)&SS& -0.50 (-0.41$^d$,-0.49$^e$-0.51$^f$)& 
$0.23$&$800\pm 50$\\
Au(001)&1& -0.66 (-0.69$^g$,-0.63$^h$)&$1.05$&$88\pm 4$\\
       &2& -0.18&$1.05$&$20\pm 2$\\
\mr
\multicolumn{2}{l}
{$^a$ See Ref.~\cite{link01}}
&&
\multicolumn{2}{l}
{$^e$ See Ref.~\cite{nicolay01}}
\\
\multicolumn{2}{l}
{$^b$ See Ref.~\cite{kinoshita96}}
&&
\multicolumn{2}{l}
{$^f$ See Ref.~\cite{kliewer00}}
\\
\multicolumn{2}{l}
{$^c$ See Ref.~\cite{drube88}}
&&
\multicolumn{2}{l}
{$^g$ See Ref.~\cite{ciccacci94}}
\\
\multicolumn{2}{l}
{$^d$ See Ref.~\cite{lashell96}}
&&
\multicolumn{2}{l}
{$^h$ See Ref.~\cite{straub84}}
\\
\end{tabular}
\end{indented}
\end{table}

The results in table \ref{table:results} indicate that the spin-orbit splitting of
$n=1$ image states at Ir, Pt and Au surfaces is an order of magnitude smaller than that of 
the Au(111) Shockley surface
state, and that of the $n=2$ states smaller still. These trends reflect the differing 
extents to which the corresponding wave functions penetrate the crystal and experience
the spin-orbit interaction at the ion cores. At the (001) surfaces 
$\Delta E_n^{\mathrm{Ir}}<\Delta E_n^{\mathrm{Pt}}<\Delta E_n^{\mathrm{Au}}$ which might be expected
given the increasing atomic number ($Z^{\mathrm{Ir}}=77,
Z^{\mathrm{Pt}}=78, Z^{\mathrm{Au}}=79$),
but at the (111) surface $\Delta E_n^{\mathrm{Ir}}>\Delta E_n^{\mathrm{Pt}}$, pointing to
a more complicated ``band-structure'' effect. In figure \ref{fig:splits} the splitting 
of the Au(001) $n=1$ image state is also seen to be affected as it disperses towards the
band edge of continuum levels. We return to this later.

\section{Phase accumulation model}
\label{section:phase}

\begin{figure}
\epsfxsize= 80mm
\centerline{\epsffile{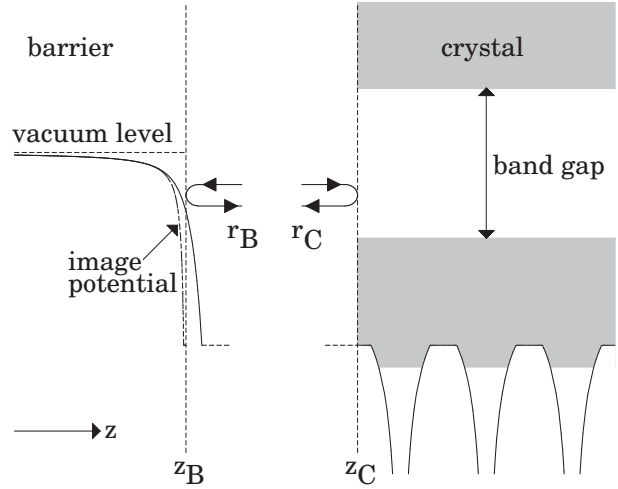}}
\caption{Schematic illustration of quantities entering the phase accumulation model.}
\label{fig:phase}
\end{figure}

The ``standard model'' for understanding image state energies is the phase accumulation 
model \cite{echenique78} in which surface states 
are envisaged as one-dimensional waves trapped by multiple reflection from 
the surface barrier and the crystal. 
In a region of constant potential (see figure \ref{fig:phase})
between barrier and crystal (which may be infinitesimal in width) the electron 
wave function can be expressed in terms of forward and backward travelling 
plane waves
\begin{equation}
\psi(z)=A\exp(\rmi kz)+B\exp(-\rmi kz).
\label{eqn:wave}
\end{equation}
The two components are related at the barrier reference plane $z=z_{\rm B}$ 
by the barrier 
reflection coefficient $r_{\rm B}$,
\(
\psi(z)\propto \exp(-\rmi k(z-z_{\rm B}))+r_{\rm B}\exp(\rmi k(z-z_{\rm B}))
\)
and at the crystal reference plane $z=z_{\rm C}$ 
by the crystal reflection coefficient $r_{\rm C}$,
\(
\psi(z)\propto \exp(\rmi k(z-z_{\rm C}))+r_{\rm C}\exp(-\rmi k(z-z_{\rm C}))
\)
which together with (\ref{eqn:wave}) give rise to the condition for a surface state to exist:
\begin{equation}
r_{\rm B}r_{\rm C}\exp(2\rmi kd)-1=0, \qquad d=z_{\rm B}-z_{\rm C}.
\end{equation}
For energies below the vacuum level and coincident with
the crystal band gap the reflection probability at both crystal and barrier
are unity and the reflection coefficients may be written in terms of phases:
$r_{\rm B}=\exp (\rmi\phi_{\rm B})$, $r_{\rm C}=\exp(\rmi\phi_{\rm C})$. The 
surface state condition then becomes
\begin{equation}
\phi_{\rm B}+\phi_{\rm C}+2kd=2\pi n \quad n=0,1,\dots
\label{eqn:phase}
\end{equation}
which is a condition on the round-trip phase accumulated by the electron wave.
The phases in (\ref{eqn:phase}) increase with energy.
The crystal phase
increases from 0 to $\pi$ as the energy sweeps across the band gap,
changing most rapidly near the band edges. Towards the bottom
of the gap it is this variation in $\phi_{\rm C}$ which will determine whether or not
(\ref{eqn:phase}) is satisfied, so that any surface state that does
arise is usually referred to as ``crystal-derived''. 
On the other hand, $\phi_{\rm B}$ increases more and more rapidly as  
the energy approaches the vacuum energy, 
varying to a good approximation as
\begin{equation}
\phi_{\rm B}(E)=\pi\left(\sqrt{\frac{3.4~\mathrm{eV}}{-E}}-1\right).
\label{eqn:barphase}
\end{equation}
In combination with (\ref{eqn:phase}) this yields a Rydberg-like
series of image states
\begin{equation}
E_n=\frac{-0.85~\mathrm{eV}}{(n+a)^2}, \quad n=1,2,\dots
\label{eqn:ryd}
\end{equation}
where the quantum defect $a=(1-[\phi_{\rm C}+2kd]/\pi)/2$ 
may usually be considered constant over the range of energies at which
the image states are found.

We now consider the extension of this model to include the spin orbit 
interaction. Introducing the electron spin in to the wavefunction in
(\ref{eqn:wave})
\begin{equation}
\psi(z)=
{\psi_\uparrow(z) \choose \psi_\downarrow(z)}=
{A_\uparrow \choose A_\downarrow}\exp(\rmi kz)+
{B_\uparrow \choose B_\downarrow}\exp(-\rmi kz),
\label{eqn:wave2}
\end{equation}
and reflection from the crystal is now described by a matrix
\begin{equation}
R_{\rm C}=
\left(
\begin{array}{cc}r_{\rm C}^{\uparrow\uparrow} & r_{\rm C}^{\uparrow\downarrow} \\ 
r_{\rm C}^{\downarrow\uparrow} & r_{\rm C}^{\downarrow\downarrow} \end{array}
\right)
\label{eqn:rc}
\end{equation}
allowing for the possibility of spin-flip upon reflection. 
With a similar matrix used to describe scattering from the barrier, the
condition for a surface state becomes
\begin{equation}
\det\left[R_{\rm B}R_{\rm C}\exp(2\rmi kd)-1\right]=0.
\label{eqn:det}
\end{equation}

The four reflection coefficients in (\ref{eqn:rc})
are not independent -- for example flux conservation requires
that $|r_{\rm C}^{\uparrow\uparrow}|^2+|r_{\rm C}^{\downarrow\uparrow}|^2=1$ within 
a gap, and for a non-magnetic crystal 
$r_{\rm C}^{\uparrow\uparrow}=r_{\rm C}^{\downarrow\downarrow}$.
In the non-magnetic case and for a planar potential 
$V=V(z)$ the electron wave functions
$\Psi_{\bi K}(\bi{r})=\psi_{\bi{K}}(z)\exp(\rmi\bi{K}\cdot\bi{r}_\| )$ are
found from the Hamiltonian 
\begin{equation}
H=-\frac{\hbar^2}{2m}\nabla^2+V+
\frac{\hbar^2}{4m^2c^2}\bsigma\cdot\left(\bnabla V_{\textstyle\times}\bi{K}\right).
\end{equation}
which may be diagonalised by a rotation in spin space
\begin{equation}
H'=U_\vartheta H U_\vartheta^{-1}=-\frac{\hbar^2}{2m}\nabla^2+V+
\frac{\hbar^2 K}{4m^2c^2} \frac{dV}{dz} \sigma_z
\label{eqn:diag}
\end{equation}
with
\begin{equation}
U_\vartheta=\frac{1}{\sqrt{2}}\left(
\begin{array}{cc}1 & -\rmi \exp(-\rmi \vartheta) \\ 1 & +\rmi \exp(-\rmi\vartheta)\end{array}
\right)
\end{equation}
where $\vartheta$ is the angle of the electron wave vector.
The Hamiltonian $H'$ does not mix spin-up and spin-down channels and so 
in this representation the reflection matrix describing scattering 
from the crystal
is diagonal:
\begin{equation}
R'_{\rm C}=\left(\begin{array}{cc}r^+_{\rm C} & 0 \\ 0 & r^-_{\rm C} \end{array}\right).
\end{equation}
Since the spin-orbit interaction is negligible in the barrier the 
barrier reflection matrix is also diagonal, and for a non-magnetic surface
$R_{\rm B}^{\uparrow\uparrow}=R_{\rm B}^{\downarrow\downarrow}=
r^+_{\rm B}=r^-_{\rm B}=r_{\rm B}$
so that the surface state condition (\ref{eqn:det}) becomes
$r^\pm_{\rm C} r_{\rm B} \exp(2\rmi kd)=1$
leading to the round-trip phase condition
\begin{equation}
\phi_{\rm C}\pm\eta+\phi_{\rm B}+2kd=2\pi n \qquad n=1,2,\dots
\label{eqn:phase2}
\end{equation}
where we have introduced $r^\pm_{\rm C}=\exp(\rmi(\phi_{\rm C}\pm\eta))$ appropriate to 
energies within a gap. The surface states now come in spin-split pairs, 
with spin orientations that may be
deduced from the spinors that are obtained by rotating back in to the 
original reference frame the spin-up and -down eigenspinors 
$\chi^{\prime +}={1 \choose 0},\chi^{\prime -}={0 \choose 1}$ 
of the primed frame:
\begin{equation}
\chi^\pm_{\vartheta}=U_\vartheta^{-1}\chi^{\prime\pm}=
\frac{1}{\sqrt{2}}
{1 \choose \pm \rmi \exp(\rmi\vartheta)}.
\end{equation}
Thus ${\widehat{\bi S}}=\pm(-\sin\vartheta,\cos\vartheta,0)=\pm 
\widehat{z}_{\textstyle\times}\widehat{K}$ and
we find that the spins lie in the surface plane 
and perpendicular to ${\bi K}$, to the left (right) for $+$ ($-$).

\begin{figure}
\centerline{\epsfxsize=160mm\epsffile{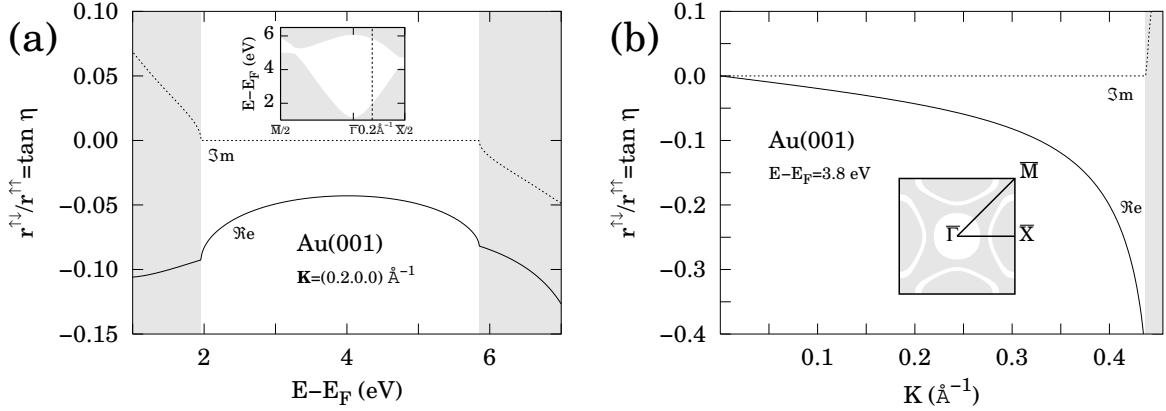}}
\caption{Variation of the spin-flip to non-spin-flip scattering ratio
$r^{\uparrow\downarrow}/r^{\uparrow\uparrow}$ for Au(001) in the major band
gap of the projected band structure. (a) as a function of energy at 
$\bi{K}=(0.2,0.0)$~\AA$^{-1}$. (b) as a function of wave vector 
along $\overline{\Gamma}\, \overline{X}$ at the mid-gap energy. The 
inset shows the projected gaps across the Brillouin zone at the same energy.}
\label{fig:tan}
\end{figure}

Rotating back to the original spin frame gives the reflection matrix 
(\ref{eqn:rc}) as
\begin{equation}
R_{\rm C}=U_\vartheta^{-1} R'_{\rm C} U_\vartheta
=\exp(\rmi\phi_{\rm C})\left(\begin{array}{cc}
\cos \eta & \exp(-\rmi\vartheta)\sin\eta \\
-\exp(\rmi\vartheta)\sin\eta & \cos \eta
\end{array} \right).
\label{eqn:Rc}
\end{equation}
In figure \ref{fig:tan} we illustrate the variation in $\tan\eta$ at the Au(001) surface 
calculated from the reflection matrix found using the relativistic multiple-scattering 
method of section \ref{section:rlkkr}.
For $\bi{K}$ along $\overline{\Gamma}\, \overline{X}$  
(see inset in figure \ref{fig:tan}) equation (\ref{eqn:Rc})
$\tan\eta=r^{\uparrow\downarrow}/r^{\uparrow\uparrow}$. The relativistic multiple-scattering
calculations include the atomic structure of the surface and the 
crystal potential is not one-dimensional, but the reflection coefficient 
for specular reflection behaves in a very similar manner to that of a 
one-dimensional crystal, especially for small
$K$ where the wave function varies only slowly across the surface.
In particular we find that the angular variation predicted by equation (\ref{eqn:Rc}) is 
satisfied to within a percent or so. It is evident from figure \ref{fig:tan} that the 
magnitude of the spin-orbit induced phase change is small, 
and away from the band edges $\eta$ is approximately 
independent of energy at Au(001) and linear in $K$:
$\eta^{\rm{Au}}_{(001)}(E,K)\simeq [-0.25~\mathrm{\AA}] K$.

\begin{figure}
\epsfxsize= 50mm
\centerline{\epsffile{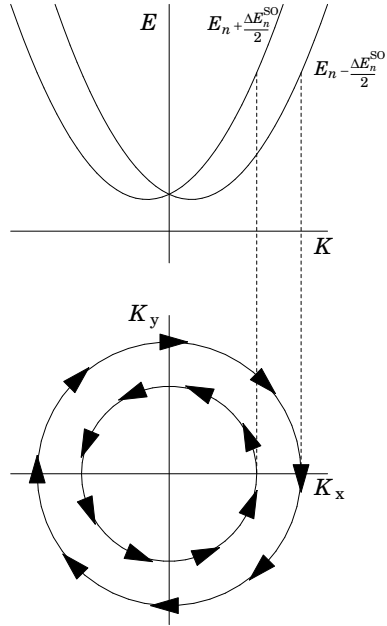}}
\caption{Spin orientation of spin-orbit split surface states for $\eta<0$.}
\label{fig:spin}
\end{figure}

To first order the round-trip phase condition 
(\ref{eqn:phase2}) is satisfied at energies $E_n\pm\Delta E_n^{\rm SO}/2$
where
\begin{equation}
\Delta E_n^{\rm SO}\simeq -2\eta
\left(\left. \frac{\rmd}{\rmd E}(\phi_{\rm B}+\phi_{\rm C}+2kd)\right|_{E_n}\right)^{-1},
\label{eqn:energies}
\end{equation}
neglecting the energy dependence of $\eta$ which is small compared 
to the other phases.
For all the surfaces that we have studies we have found that $\eta$ does not change
sign across the band gap, and the denominator in (\ref{eqn:energies}) is positive. 
Hence the model predicts a series of spin-orbit split states with 
identical spin orderings, which we have confirmed is also the case in the 
relativistic multiple scattering calculations described in section \ref{section:rlkkr}.
In particular with $\eta<0$ the surface states that exist 
are split with the lower of each pair of levels having the spin pointing 
to the right of 
$\bi{K}$, as shown in figure \ref{fig:spin}. This is in agreement with the 
spin assignments shown in Henk {\it et al.} \cite{henk03} for the Au(111) surface state, 
but disagrees with those given in Ref. \cite{nicolay01}.

When the gap contains image states, over the narrow range of energies within 
which the image states are found the energy-dependence of the round-trip phase
is dominated by the variation in the barrier phase (\ref{eqn:barphase}) and 
then
\begin{equation}
\Delta E_n^{\rm SO}\simeq-2\eta
\left(\left. \frac{\rmd \phi_{\rm B}}{\rmd E}\right|_{E_n}\right)^{-1}=
-\frac{\eta\times 1.7~\mathrm{eV}}{\pi (n+a)^3}
\label{eqn:de}
\end{equation}
Thus the spin-splittings exhibit the same scaling as the lifetime broadening
\cite{echenique78}, in each case the behaviour ultimately originating in the 
variation with $n$ of the wave function overlap with the crystal. 
We also see from
(\ref{eqn:de}) that the linear-in-$K$ behaviour of $\Delta E_n^{\rm SO}$ arises from
similar behaviour in $\eta$.
Since $\eta$ is small $\tan\eta\simeq\sin\eta\simeq\eta$
to a good approximation, and hence $\eta\simeq-|r^{\uparrow\downarrow}|$. 
Thus the spin-orbit splitting is directly related to the spin-flip
scattering rate, which could therefore be determined from experimental 
values of image state splittings.
At Au(001), $E_1\simeq-0.66$ meV (table \ref{table:results}), so combining
equations (\ref{eqn:ryd}) and (\ref{eqn:de}) gives
\begin{equation}
\Delta E_1^{\rm SO}\simeq -\eta \times \frac{1.7~\mathrm{eV}}{\pi}
\left(\frac{0.66}{0.85}\right)^{3/2}
=-\eta\times 0.37~\mathrm{eV}
\end{equation}
and a splitting of 22 meV (figure \ref{fig:splits}) at $K=0.25$ \AA$^{-1}$ 
yields $\eta\simeq 0.06$, which agrees with the value found from the 
multiple-scattering calculations shown in figure \ref{fig:tan}. 
Finally, we note that equation (\ref{eqn:de}) will not hold
near band edges where the energy-dependence of the crystal phase $\phi_{\rm C}$ cannot be
neglected. This is the origin of the anomalous dispersion shown for the $n=1$ state 
at Au(001) in figure \ref{fig:splits}.

\section{Discussion}

To summarise, we have investigated the effect of the spin-orbit interaction on image state
electrons at the (111) and (001) surfaces of Ir, Pt, and Au. Non-perturbative calculations 
that use relativistic multiple-scattering theory with self-consistent potentials and a 
parameterised surface barrier predict Rashba-like dispersion of the image state bands
with splittings for $n=1$ that are a factor 10-20 times smaller than that of the 
Au(111) Shockley state, for which our results are in good agreement with experiment 
and previous theory. Extending
the phase accumulation model to include spin-orbit scattering, we find that the splittings
scale as $1/(n+a)^3$, where $a$ is the quantum defect, and are 
directly related to the spin-flip scattering rate at the surface.
The largest image state splittings we find are at Au(001) for which 
$\Delta E_1^{\rm SO}=22$ meV at $K=0.25$ \AA$^{-1}$.
This is larger than some exchange splittings of image states that have previously been 
resolved (e.g. $18\pm 13$ meV at Ni(111) \cite{passak92} and $13\pm 13$ meV at Ni(001)
\cite{starke92}) by exploiting the spin resolution available in spin-resolved inverse 
photoemission, suggesting that the spin-orbit splitting may also be observable 
with an appropriate experimental set-up.
The splittings are comparable to lifetime broadenings of late-transition and 
noble metal image states \cite{echenique00,rhie03}, indicating that account of spin-orbit 
effects may be necessary when determining image state lifetimes from lineshape analysis 
for 5$d$ metals.

\end{document}